\documentclass[lettersize,journal]{IEEEtran}
\usepackage{amsmath,amsfonts}
\usepackage{algorithmic}
\usepackage{algorithm}
\usepackage{array}
\usepackage{textcomp}
\usepackage{stfloats}
\usepackage{url}
\usepackage{verbatim}
\usepackage{graphicx}
\usepackage{microtype}
\usepackage{cite}
\usepackage{siunitx}
\usepackage[utf8]{inputenc}
\PassOptionsToPackage{usenames,dvipsnames}{xcolor}
\usepackage{booktabs}
\usepackage{csquotes}
\usepackage{adjustbox}
\usepackage[flushleft]{threeparttable}
\usepackage{hyperref}
\usepackage{orcidlink}
\usepackage{multirow}

\usepackage[skip=0pt]{caption}
\usepackage{subcaption}
\usepackage{tikz}
\usetikzlibrary{shapes.geometric, arrows.meta, positioning, calc}
\usetikzlibrary{fit}
\usepackage{pgfplots}
\usepackage{pgfplotstable}
\usepgfplotslibrary{groupplots}
\pgfplotsset{compat=1.18}
\usepackage[siunitx, american]{circuitikz}
\ctikzset{
  transistors/scale=1,
  resistors/scale=0.7,
  capacitors/scale=0.7,
  bipoles/length=20pt,    
}

\usepackage{xfrac}
\usepackage{xparse}
\ExplSyntaxOn
\DeclareExpandableDocumentCommand{\convertlen}{ O{cm} m }
{
	\dim_to_decimal_in_unit:nn { #2 } { 1 #1 } cm
}
\ExplSyntaxOff

\usepackage{xspace}
\xspaceaddexceptions{)}
\usepackage[acronyms]{glossaries}
\hypersetup{hidelinks}

\newglossaryentry{vdd}{
	name={\ensuremath{V_{\text{DD}}}},
	description={supply voltage},
	symbol=Vdd,
	sort=vdd}
\newglossaryentry{vth}{
	name={\ensuremath{V_{\text{TH}}}},
	description={threshold voltage},
	symbol=Vth,
	sort=vth}
\newglossaryentry{gnd}{
	name={GND},
	description={El. ground},
	symbol=GND}

\newacronym{aef}{AEF}{average electric field}
\newacronym{dnn}{DNN}{deep neural network}
\newacronym{nn}{NN}{Neural network}
\newacronym{kan}{KAN}{Kolmogorov-Arnold network}
\newacronym{fkan}{FKAN}{Fourier KAN}
\newacronym{kart}{KART}{Kolmogorov-Arnold representation theorem}
\newacronym{cpn}{CPN}{current/charge prediction network}
\newacronym{pgn}{PGN}{parameter generator network}
\newacronym{gnn}{GNN}{graph neural network}
\newacronym{rnn}{RNN}{recurrent neural network}
\newacronym{cnn}{CNN}{convolutional neural network}
\newacronym{aspp}{ASPP}{atrous spatial pyramid pooling}
\newacronym{mape}{MAPE}{mean absolute percentage error}
\newacronym{rmse}{RMSE}{root mean square error}
\newacronym{dtco}{DTCO}{design technology co-optimization}
\newacronym{sta}{STA}{static timing analysis}
\newacronym{iou}{IoU}{intersection over union}
\newacronym{fom}{FoM}{figure of merit}
\newacronym{miou}{MIoU}{mean intersection over union}
\newacronym{espp}{EsPP}{electrostatic potential profile}
\newacronym{mse}{MSE}{mean square error}
\newacronym{bjt}{BJT}{bipolar junction transistor}
\newacronym{ml}{ML}{Machine learning}
\newacronym{mlp}{MLP}{multi-layer perceptron}

\newacronym{rnet}{R-Net}{resistor network}

\newacronym{tcad}{TCAD}{Technology CAD}
\newacronym{cad}{CAD}{computer-aided design}
\newacronym{spice}{SPICE}{Simulation Program with Integrated Circuit Emphasis}
\newacronym{hspice}{HSPICE}{PrimeSim HSPICE}
\newacronym{ddic}{DDIC}{display driver integration circuit}
\newacronym{ss}{SS}{sub-threshold slope}

\newacronym{fet}{FET}{Field Effect Transistor}
\newacronym{fdsoi}{FDSOI}{Fully-Depleted Silicon on Insulator}
\newacronym{finfet}{FinFET}{fin field-effect transistor}
\newacronym{fe}{FE}{Ferroelectric}
\newacronym{fefet}{FeFET}{Ferroelectric FET}
\newacronym{mlc}{MLC}{multi-level cell}
\newacronym{nvm}{NVM}{non-volatile memory}

\newacronym{mac}{MAC}{multiply-accumulate}
\newacronym{adc}{ADC}{analog-digital-converter}

\newacronym{dut}{DUT}{device under test}
\newacronym{smu}{SMU}{source measurement unit}

\newacronym{wl}{WL}{wordline}
\newacronym{bl}{BL}{bitline}
\newacronym{sl}{SL}{signal line}

\newacronym{nls}{NLS}{nucleation-limited switching}
\newacronym{ds}{DS}{Domain Stochasticity}

\newacronym{dtd}{DTD}{Device-to-Device}
\newacronym{ctc}{CTC}{Cycle-to-Cycle}

\newacronym{stt}{STT}{Spin Transfer Torque}
\newacronym{rram}{ReRAM}{Resistive RAM}
\newacronym{pde}{PDE}{partial differential equation}

\newcommand{\nn}{\gls{nn}\xspace}

\newcommand{\mape}{\gls{mape}\xspace}

\newcommand{\fe}{\gls{fe}\xspace}
\newcommand{\fefet}{\gls{fefet}\xspace}
\newcommand{\fefets}{\glspl{fefet}\xspace}
\newcommand{\mlc}{\gls{mlc}\xspace}

\newcommand{\adc}{\gls{adc}\xspace}

\newcommand{\bitline}{\gls{bl}\xspace}
\newcommand{\wordline}{\gls{wl}\xspace}
\newcommand{\signalline}{\gls{sl}\xspace}

\newacronym{mw}{MW}{memory window}

\newcommand{\vth}{\gls{vth}\xspace}
\newcommand{\vds}{\ensuremath{V_{\text{DS}}}\xspace}
\newcommand{\vgs}{\ensuremath{V_{\text{GS}}}\xspace}
\newcommand{\vs}{\ensuremath{V_{\text{S}}}\xspace}

\newcommand{\gm}{\ensuremath{g_{\text{m}}}\xspace}
\newcommand{\gds}{\ensuremath{g_{\text{DS}}}\xspace}

\newcommand{\vwrite}{\ensuremath{V_{\text{W}}}\xspace}
\newcommand{\vgd}{\ensuremath{V_{\text{GD}}}\xspace}
\newcommand{\vpol}{\ensuremath{V_{\text{P}}}\xspace}
\newcommand{\vread}{\ensuremath{V_{\text{R}}}\xspace}

\newcommand{\smus}{\glspl{smu}\xspace}

\newcommand{\nvm}{\gls{nvm}\xspace}

\newcommand{\lvt}{low-\vth}
\newcommand{\hvt}{high-\vth}

\newcommand{\ids}{\ensuremath{I_{\text{DS}}}\xspace}
\newcommand{\idvg}{\ensuremath{\ids\text{-}\vgs}\xspace}
\newcommand{\idvd}{\ensuremath{\ids\text{-}\vds}\xspace}

\newcommand{\grid}[1]{\ensuremath{#1\!\times\!#1}\xspace}
\newcommand{\nmgrid}[2]{\ensuremath{#1\!\times\!#2}\xspace}

\newcommand{\code}[1]{\texttt{#1}}
\newcommand{\logicone}{\code{1}\xspace}
\newcommand{\logiczero}{\code{0}\xspace}

\newcommand{\hfo}{HfO$_{2}$\xspace}

\def\nobreakbefore{%
  \relax\ifvmode\else
    \ifhmode
      \ifdim\lastskip > 0pt\relax
        \unskip\nobreakspace
      \fi
    \fi
  \fi
}
\let\oldcite\cite
\renewcommand\cite{\nobreakbefore\oldcite}

\definecolor{fefetcolor}{RGB}{255, 183, 3}

\usepackage{cleveref}

\crefname{enumi}{Step}{Steps}
\crefname{section}{Section}{Sections}
\crefname{subsection}{Section}{Sections}
\crefname{figure}{Figure}{Figures}
\crefname{algocf}{Algorithm}{Algorithms}
\crefname{algorithm}{Algorithm}{Algorithms}
\crefname{algocf}{Algorithm}{Algorithms}
\crefname{table}{Table}{Tables}
\crefname{equation}{Equation}{Equations}
\crefname{eqnarray}{Equation}{Equations}
\crefname{appendix}{Appendix}{Appendix}

\begin{document}

\title{NN-Based FeFET Compact Modeling for SPICE Simulation}

\author{Rodion~Novkin$^{\orcidlink{0009-0006-6632-9804}}$, Simon~Thomann$^{\orcidlink{0000-0002-7902-9353}}$,~\IEEEmembership{Member,~IEEE}, and Hussam~Amrouch$^{\orcidlink{0000-0002-5649-3102}}$,~\IEEEmembership{Member,~IEEE}%
\thanks{This work was supported by the German Research Foundation (DFG) Device-Circuit Co-Optimization for Reliable FeFET-based Brain-inspired Computing in the Face of Variability (DeCOP) under Grant AM 534/8-1 and Grant 547377347. The project on which this report is based was sponsored by the Federal Ministry of Research, Technology and Space under the funding code SWC3.0 16IS23069 \enquote{RAVEN}. Responsibility for the content of this publication lies with the authors. \emph{(Corresponding author: Hussam Amrouch)}}
\thanks{
Rodion Novkin, Simon Thomann, and Hussam Amrouch are with the Technical University of Munich; TUM School of Computation, Information and Technology; Chair of AI Processor Design; Munich Institute of Robotics and Machine Intelligence, 80333 Munich, Germany (e-mail: \nolinkurl{{rodion.novkin,s.thomann,amrouch}@tum.de}). 
}

}


\markboth{}
{Shell \MakeLowercase{\textit{Novkin et al.}}: Derivative-Free Neural Network Training with Measurement Data for FeFET Compact Modeling}


\maketitle

\begin{abstract}
Prototyping and early-stage circuit evaluation for emerging transistor technologies require fast and accurate compact models for SPICE simulations. Standard physics-based compact models, such as BSIM, offer high fidelity but demand extensive domain expertise and time-consuming equation formulation for every new physical effect, significantly slowing technology development. \nn-based models offer a scalable alternative approach, but they rely heavily on smooth numerical solver data with explicit high-order derivative information. However, measurement data suffers from complex noise patterns, making derivative extraction challenging and complicating \nn-based transistor modeling. This work presents a derivative-free \nn training framework employing spectral normalization to naturally smooth derivative behavior and filter measurement noise. Using drain current measurements from a fabricated \qty{28}{nm} ferroelectric device, we design an \nn-based compact model that captures the full spectrum of polarization amplitudes and supports dynamic polarization switching in transient simulations. To demonstrate its capability, we deploy the model in a commercial SPICE software to simulate write and read operations for large-scale memory arrays with up to \grid{512} size, including parasitic effects. Our approach bridges raw measurements with \nn-based SPICE-compatible transistor modeling, significantly reducing model development time and enabling rapid circuit-level prototyping for emerging devices.
\end{abstract}


\section{Introduction}
\glsresetall
\IEEEPARstart{C}{ircuit} simulations in the rapidly evolving semiconductor industry demand accurate and fast transistor models. Berkeley short-channel IGFET (BSIM) \cite{Chauhan14} covers a wide range of modern transistor technologies and remains the most commonly used industry-standard model for analog and digital circuit designs. BSIM relies on physics-based equations to accurately capture underlying transistor behavior, enabling the model to achieve excellent performance in modeling device characteristics. However, developing an analytical solution to accurately describe the underlying physics of device operation is often time-consuming and requires substantial domain expertise, which significantly slows down the development process for emerging technologies since all new physical effects need to be properly studied, characterized with equations, and implemented in the model. \nn-based transistor modeling received significant attention over the recent years as a promising alternative to conventional models. NNs can effectively create a standalone, compact model for specific working conditions dictated by the range of data and input parameters used for NN training. Matrix multiplication, prevalent in many \nn architectures, is a simple operation that enables \nn-based models to significantly reduce model complexity and accelerate SPICE simulations.

While \nn-based transistor models have demonstrated success across various devices \cite{Tung2023, Wang21, Eom24, NovkinDualNN25}, prior implementations largely relied on synthetic data from numerical solvers (e.g., SPICE tools) that provide high-precision derivative information. However, experimental measurement data is inherently noisy, and in addition to standard Gaussian noise, there are also various anisotropic, conditionally dependent, or structured noise patterns, making the extraction of derivative information with conventional numerical and filtering methods challenging. Since derivative smoothness and stability are essential for SPICE solver convergence, fitting continuous, noise-resilient NN models to measured data remains a significant hurdle. Therefore, a reliable way to handle measurement data with \nn-based transistor models has yet to be found.

\noindent
\textbf{Our main contributions within this paper are as follows:}

        \noindent \textbf{(1)} We propose a derivative-free approach for \nn-based transistor model training using spectral normalization, which guides \nn derivative behavior and filters noise present in the measurement data.
        
        \noindent
        \textbf{(2)} We measure transfer characteristics of a \qty{28}{nm} \fefet devices and develop an accurate \nn-based transistor model, which covers the entire range of polarization amplitudes and supports dynamic polarization changes in transient simulations.

        \noindent
        \textbf{(3)} For comprehensive testing, we deploy the \nn-based \fefet model within a commercial SPICE tool and simulate the write and read operations across various memory array configurations. The size of arrays varies from \grid{2} up to \grid{512}, and all arrays incorporate parasitic elements.

\section{Background}
\label{sec:background}

\subsection{Ferroelectric-Based Transistors}
\label{sec:fefet_background}



\begin{figure}[t]
    \centering
    \begin{subfigure}{0.38\columnwidth}
        \footnotesize
        \centering
        \begin{tikzpicture}[xscale=0.45, yscale=0.6]
    \draw[fill=blue!20,draw] (-5, -0.3) rectangle (1.8, 1.9) node[pos=.5] {p-Si};
    \draw[fill=red!20,draw] (-4.5, 1.1) rectangle (-3, 1.9) node[pos=.5] {n+};
    \draw[fill=red!20,draw] (-0.2, 1.1) rectangle (1.3, 1.9) node[pos=.5] {n+};
    \draw[fill=gray!30,draw] (-3, 1.9) rectangle (-0.2, 2.5) node[pos=.5] {SiO\textsubscript{2}};
    \draw[fill=fefetcolor,draw] (-3, 2.5) rectangle (-0.2, 3.5) node[pos=.5] {\hfo};
    \draw[fill=green!40,draw] (-3, 3.5) rectangle (-0.2, 4.4) node[pos=.5] {Metal};
    

    \draw[thick] (-4, 1.9) -- (-4, 2.3) coordinate (drainDot);
    \filldraw[black] (drainDot) circle (2pt);
    \node[above] at (drainDot) {\vds};

    \draw[thick] (0.75, 1.9) -- (0.75, 2.3) coordinate (sourceDot);
    \filldraw[black] (sourceDot) circle (2pt);
    \node[above] at (sourceDot) {\vs};

    \draw[thick] (-1.6, 4.4) -- (-1.6, 4.8) coordinate (gateDot);
    \filldraw[black] (gateDot) circle (2pt);
    \node[above] at (gateDot) {\vgs};
    
\end{tikzpicture}
        \caption{\acrshort{fefet} gate stack}
        \label{fig:fefet_gate_stack}
    \end{subfigure}
    \begin{subfigure}{0.6\columnwidth}
        \footnotesize
        \centering
        \input{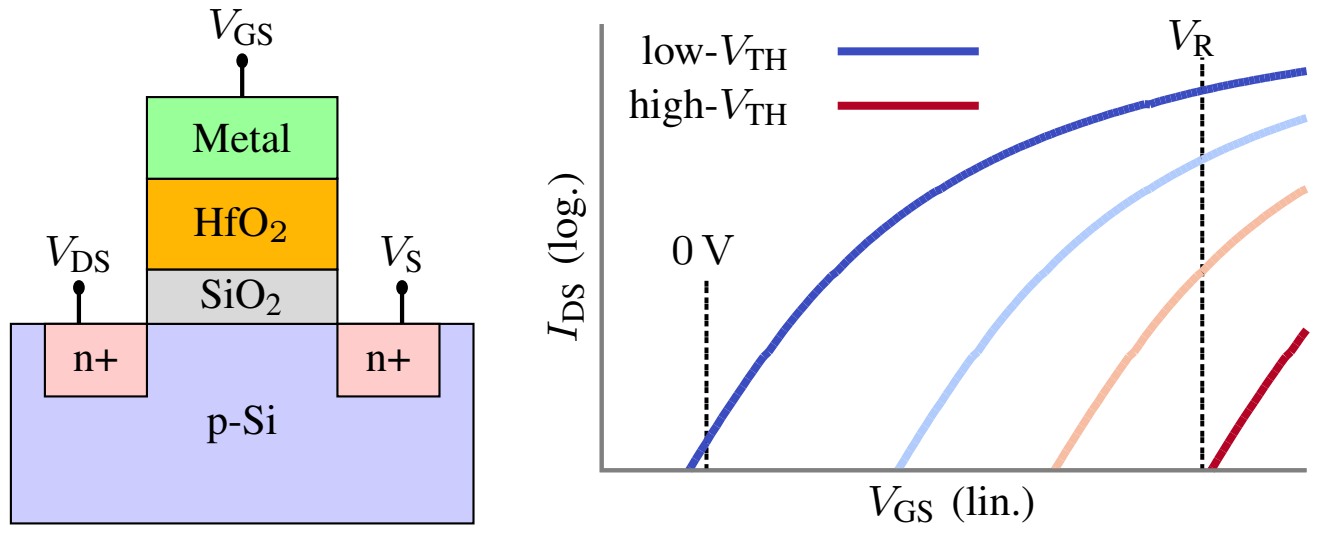}
        \vspace{-0.3cm}
        \caption{\acrshort{fefet} \idvg sketch}
        \label{fig:iv_sketch}
    \end{subfigure}
    \caption{(a) \acrshort{fefet} gate stack with the added \hfo-based \acrshort{fe} layer.
    (b) Sketch of the \idvg characteristics of \acrshort{fefet} with intermediate states.
    At $\vgs = \qty{0}{V}$ the current is OFF, even in the \lvt state, while the ON current at \vread (e.g., \qty{1}{V}) depends on the polarization state.}  
    \label{fig:fefet_background}
\end{figure}

Adding a thin (e.g., \qty{10}{nm}) \hfo \fe layer to a MOSFET gate stack yields a CMOS-compatible \fefet \cite{boeschke2011ferro_hfo,boeschke2011first_fefet} (see \cref{fig:fefet_gate_stack}). The electric polarization of the \fe layer interacts with the channel below, repelling or attracting carriers depending on the polarization direction. Since polarization of the \fe layer is permanent, it modulates channel carriers to form two distinct threshold voltage (\vth) states for \nvm storage (see \cref{fig:iv_sketch}). Negative polarization maximizes \vth (\hvt state associated with \enquote{\logiczero}), while positive polarization minimizes it (\lvt state associated with \enquote{\logicone}). Gate pulses exceeding the coercive field (e.g., \qty{\pm 4}{V}) switch the polarization state by creating the necessary electric field across the \fe layer, which is stronger than the coercive field. Additionally, modulating the pulse amplitude or width achieves partial polarization, shifting \vth to intermediate states for \mlc operation \cite{jerry2017fefet_synapse}.

\subsection{FeFET-Based Memory Arrays}
\label{sec:memory_array_background}


\def\gridSize{3} 

\begin{figure*}[t]
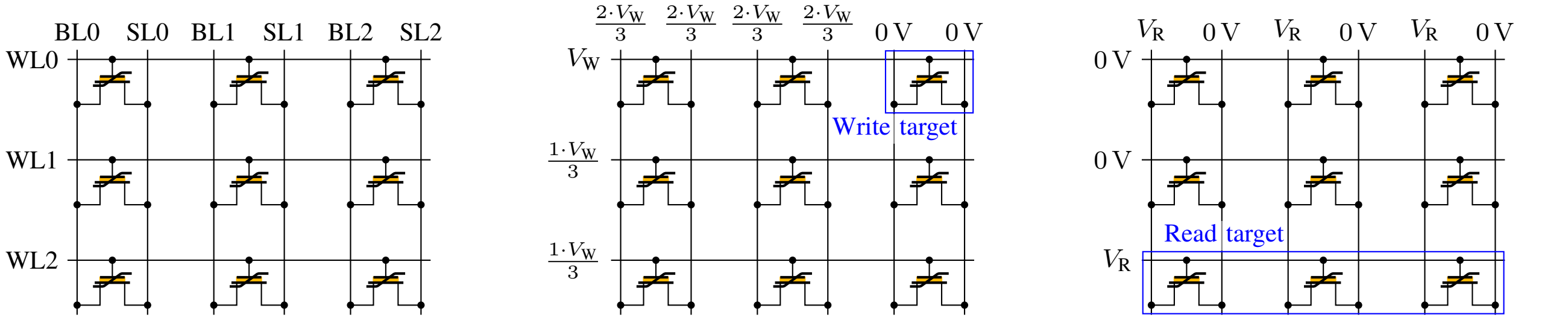

    \centering
    \begin{subfigure}{0.32\linewidth}
        \footnotesize
        \centering
        \begin{tikzpicture}

\input{1_initial/figures/memory_template}

\foreach \i in {0,...,\maxIdx}
    \draw (bl\i) node[above]{\acrshort{bl}\i}
          (sl\i) node[above]{\acrshort{sl}\i}
          (wl\i) node[left]{\acrshort{wl}\i}
            ;

\end{tikzpicture}
        \caption{\fefet memory array architecture}
        \label{fig:memory_architecture}
    \end{subfigure}
    \begin{subfigure}{0.32\linewidth}
        \footnotesize
        \centering
        \begin{tikzpicture}

\input{1_initial/figures/memory_template}

\foreach \i in {1,...,\maxIdx}
    \draw 
        (wl\i) node[left]{$\frac{1\cdot\vwrite}{3}$}
        ;
        
\pgfmathtruncatemacro{\writeMaxIdx}{\gridSize-2}
\foreach \i in {0,...,\writeMaxIdx}
    \draw 
        (sl\i) node[above]{$\frac{2\cdot\vwrite}{3}$}
        (bl\i) node[above]{$\frac{2\cdot\vwrite}{3}$}
        ;

\draw 
    (wl0) node[left]{\vwrite}
    (sl\maxIdx) node[above]{\qty{0}{V}}
    (bl\maxIdx) node[above]{\qty{0}{V}}
    ;

\node [draw, blue, fit=(mn\maxIdx-\maxIdx.G) (mn\maxIdx-\maxIdx.D) (mn\maxIdx-\maxIdx.S)] (writeTarget){};
\node [below, blue, fill=white,fill opacity=0.5, text opacity=1, inner sep=1] at ([xshift=0.11cm]writeTarget.south west) {\footnotesize Write target};

\end{tikzpicture}
        \caption{\fefet memory array write scheme}
        \label{fig:memory_write_scheme}
    \end{subfigure}
    \begin{subfigure}{0.32\linewidth}
        \footnotesize
        \centering
        \begin{tikzpicture}

\input{1_initial/figures/memory_template}

\foreach \i in {0,...,\maxIdx}
    \draw 
        (bl\i) node[above]{\vread}
        (sl\i) node[above]{\qty{0}{V}}
        ;

\pgfmathtruncatemacro{\readMaxIdx}{\gridSize-2}
\foreach \i in {0,...,\readMaxIdx}
    \draw 
        (wl\i) node[left]{\qty{0}{V}}
        ;
        
\draw (wl\maxIdx) node[left]{\vread};

\node [draw, blue, fit=(mn\maxIdx-0.G) (mn\maxIdx-0.D) (mn0-0.S)] (readTarget){};
\node [above, blue, fill=white,fill opacity=0.5, text opacity=1, inner sep=1] at ([xshift=0.9cm]readTarget.north west) {\footnotesize Read target};

\end{tikzpicture}
        \caption{\fefet memory array read scheme}
        \label{fig:memory_read_scheme}
    \end{subfigure}
    \caption{(a) \fefet memory array architecture at the example of a \grid{\gridSize} array.
    (b) \fefet memory array write scheme, targeting the top right \fefet.
    (c) \fefet array read-out bias scheme for the last row.}  
    \label{fig:fefet_memory}
\end{figure*}

Using \fefet, dense random access memory arrays can be implemented.
Following a design from the literature, we show the architecture of the memory array in \cref{fig:memory_architecture} at the example of a \grid{\gridSize} array \cite{reis2019fefet_dense_memory}, where \fefets form a grid-like structure and serve as memory cells.

\cref{fig:memory_write_scheme} shows the write scheme to change the polarization of a specific \fefet.
It shows all biases per line to write the top-right \fefet (target), while the states of the other \fefets are not disturbed.
The targeted \fefet perceives the full potential difference of the write voltage \vwrite across the \fe layer, with its source and drain terminals grounded.
To write the value to the target \fefet while not disturbing the state of other \fefets in that row, a \sfrac{1}{3}-based disturbance-free write scheme is employed \cite{reis2019fefet_dense_memory}.
For this, \bitline and \signalline of the non-target columns are biased with $\sfrac{2}{3}\cdot \vwrite$ to reduce the strength of the electric field below the coercive field.
Similarly, the \wordline of the non-target rows needs to be biased with $\sfrac{1}{3}\cdot \vwrite$ to not disturb their states.

To read out a stored data word as shown in \cref{fig:memory_read_scheme}, a row is selected by applying \vread to a \wordline, which enables it.
By applying \vread also to the \glspl{bl}, the current per \bitline depends on the state of enabled \fefet and encodes the stored value.
Using an \adc, the \bitline currents can be converted back into digital values for further processing.

\subsection{Device Measurement}
\label{sec:device_measurements}

\begin{figure}
    \centering
    \includegraphics[width=0.9\columnwidth]{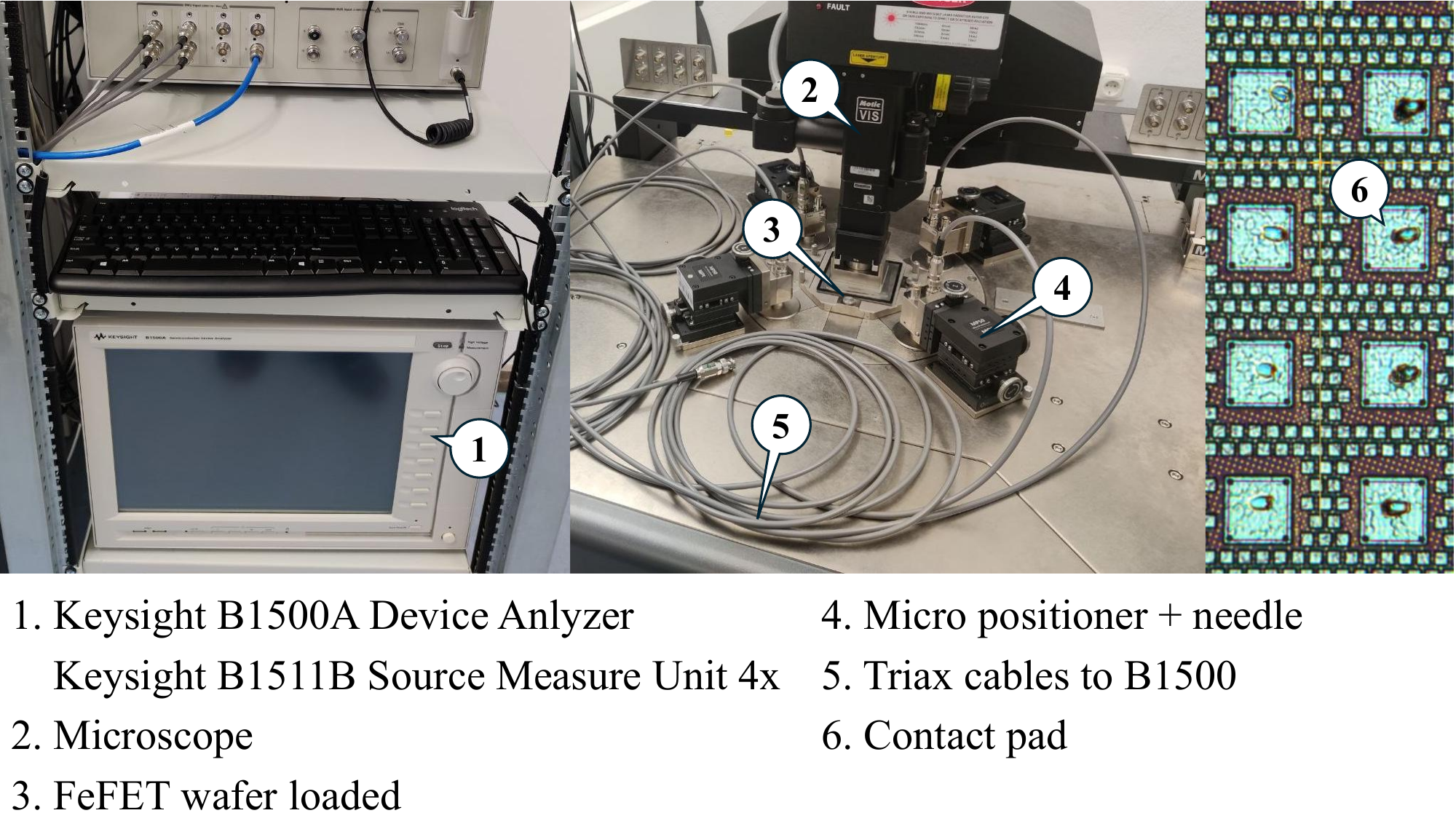}
    \caption{Measurement setup for the electrical characterization of the \fefet on \qty{300}{mm} wafer.}
    \label{fig:measurement_setup}
\end{figure}

To collect the measurement data that forms the foundation of the training dataset, we have measured a \nmgrid{\qty{7}{\micro\meter}}{\qty{10}{\micro\meter}} (\nmgrid{\text{width}}{\text{length}}) \fefet fabricated in a \qty{28}{nm} process (see \cref{fig:measurement_setup}).
Using an MPI~TS3000-SE wafer probe station, we have loaded a \qty{300}{mm} wafer onto the thermally controlled chuck, which is held at \qty{300}{K}.
For electrical characterization, we use the Keysight~B1500A semiconductor device analyzer mainframe.
To drive the biases and measure currents, the \num{4} installed Keysight~B1511B \smus are employed and connected via triaxial cables to the micro positioners, which contact the pads on the wafer \cite{mema2026fefet_pde,mema2026fefet_nvsram}.

To gather data on multiple polarization states, a two-step measurement protocol is used.
1) The \fefet is polarized to the target state.
For this, we reset the polarization to the \hvt state with a \qty{2}{s} long \qty{-4}{V} pulse.
Starting from this known state, we polarize the target intermediate state by applying the positive voltage pulse (\vwrite) of different amplitudes for \qty{2}{s}, which results in achieving a corresponding voltage polarization amplitude state ($\vpol=\vwrite$).
The polarization step concludes by waiting for \qty{10}{s} to allow accumulated traps at the channel-\fe interface to detrap.
2) After the \fefet has been polarized, the \idvg characteristics are measured.
For this, we sweep \vgs from \qtyrange{-0.2}{1.05}{V} in \qty{10}{mV} steps at a fixed \vds bias and grounded \vs.
To have more data close to $\vds = \qty{0}{V}$, we use a non-linear step size for the \vds biases.
For \qty{1}{mV} up to \qty{10}{mV} the resolution is \qty{1}{mV}, from \qtyrange{10}{100}{mV} the resolution is \qty{10}{mV}, and from \qtyrange{0.1}{1}{V} the resolution is \qty{25}{mV}.

The used amplitudes during the polarization step are \qtylist{0;1.5;1.8;2;2.3;2.5;2.7;3;3.5;4;4.5}{V}, i.e., \num{11} \idvg datasets in total.
Due to the polarization scheme described above, the \qty{0}{V} case represents the \hvt, since no positive-voltage pulse followed the reset.

\section{Our Derivative-Free NN Training Framework} \label{sec:method}
\subsection{NN Architecture and Training Algorithm}
Although \nn-assisted \fefet modeling was explored before \cite{Thomann2024, tseng25}, a SPICE-compatible \nn-based compact model was not introduced. A conventional approach for \nn-based transistor model training demands using first- and, sometimes, second-order derivative information and incorporating it into the loss function \cite{Tung2023,Eom24,Wang21,NovkinDualNN25}. However, in contrast to previous works, our training data was measured and not generated by numerical simulators. Therefore, the ground truth for clean and smooth derivatives does not exist. It is possible, in theory, to filter the data, but data filtering in three dimensions (gate and drain voltages, polarization amplitude) is notoriously hard because values in all three dimensions are interconnected and are supposed to represent the same smooth and continuous function $\ids=f(\vds, \vgs, \vpol)$. Since measurements were performed as \idvg sweeps by setting \vds and polarization amplitude \vpol constant, a measurement drift occurs in these dimensions, leading to noisy \ids-\vds and \ids-\vpol sweeps. Having axis-aligned or sweep-dependent noise characteristics poses additional challenges, since most filtering methods are designed for isotropic, independent, and identically distributed Gaussian noise, which is not suitable for our data. Therefore, we will explore a derivative-free and filtering-free approach for \nn-based transistor modeling.

Since we only have measurement data for positive \vds, we assume a symmetric device and force this symmetry with the drain current conversion function as:
\begin{equation}\label{eqn:cf1}
    \ids = \vds \cdot e^{y_I}; \ \ y_I=\text{NN}(|\vds|,\vgs,\vpol)
\end{equation}
This way, the sign of the current will be determined by the sign of the \vds, while the value of the current will be determined from its absolute value. We keep all voltage values and polarization amplitude values unchanged and do not normalize them. Additionally, we omitted data points with \ids below \qty{1e-12}{A}, as they are approaching the noise floor of the measurement devices and introduce significant uncertainty.

Having no derivatives available during training leaves us with the following form of the loss function:
\begin{equation}\label{eqn:loss1}
    L_I = a\cdot Er(\ids) + Er(y_I)
\end{equation}
where $Er$ is a L1 loss. Essentially, \nn is only trained with the margin loss to capture the logarithmic behavior of the curve, and a part that scales the drain current in a saturation region with coefficient $a$. The training starts with $a$=0 and ends with $a$=\num{3e+6} to ensure a sufficient contribution. 

While the loss function in \cref{eqn:loss1} is relatively simple due to the removal of derivatives, it lacks all terms that enable the correct derivative behavior. Our baseline \nn architecture consists of three hidden linear layers, each with 16 neurons, and a $\tanh$ activation function to limit the layer outputs. This architecture guarantees $C^\infty$-continuity by itself, but it still does not prevent \nn from giving wavy and extremely unstable derivatives even for a good \ids fit. Therefore, an efficient and reliable way to \textit{guide} \nn derivatives has to be introduced.

\subsection{Spectral Normalization} \label{sec:spectral_norm}
Originally proposed by Miyato et al. \cite{Miyato2018SpectralNF}, spectral normalization is widely used in the discriminator of Generative Adversarial Networks (GANs), enabling stable adversarial training and mitigating model instability without the need for extensive hyperparameter tuning or implementation of gradient penalties. Spectral normalization is a \textit{weight normalization} technique designed to control the Lipschitz constant of \nn layers by bounding their matrix spectral norms. It \textit{reparameterizes} the network weight matrix $W$ by dividing it by its largest singular value $\sigma$ (spectral norm):
\begin{equation}\label{eqn:sp_norm2}
    W_{\text{SN}} = \frac{W}{\sigma(W)}
\end{equation}

This transformation prevents NNs from producing disproportionately large output swings in response to out-of-distribution inputs. Therefore, spectral normalization is often used for GAN stabilization, adversarial robustness, and \nn architecture stability. In this work, we employ \textit{scaled} spectral normalization to mitigate measurement noise and prevent chaotic gradient behavior in the learned drain current data, since the desired \nn output cannot be physically achieved in the case of a regular spectral normalization. Therefore, we multiply the output of a normalized linear layer by a coefficient $\lambda$, which amounts to simple weight scaling. Even though the coefficients $\lambda^i$ for each layer $i$ can be made learnable, it is not recommended, as optimizers often take the path of least resistance, leading to high values of $\lambda^i$. Instead, coefficients are balanced manually with the minimum possible values to cover the entire range of the output. For the employed \nn with three hidden layers, $\lambda^{1,2}$=1.0 (base spectral normalization), $\lambda^{3}$=2.0, $\lambda^{\textit{OUT}}$=4.0, while the input layer is not normalized.

Since scaled spectral normalization targets \nn weights, for inference purposes, the weights of the pre-trained \nn can be simply \textit{recalculated} (fused with a scaled normalization layer), resulting in no additional runtime cost for the model.

\subsection{Linking Ferroelectric Polarization and Gate Voltage} \label{sec:linking_pol}
A physical \fefet has only four terminals: drain, gate, source, and body. A polarization amplitude state \vpol, used as \nn input, is an internal state of the \fefet which can be achieved by applying voltage pulses to transistor terminals. Therefore, a way to link polarization dynamics with standard transistor terminals has to be developed. In this paper, we employ a simplified model of the polarization behavior and assume that the change in the polarization state is driven by the potential difference \vgd between the transistor gate and drain terminals. Additionally, we implement certain logic and boundaries that occur in real physical \fefet devices:

\noindent \textbf{(1)} Positive polarization occurs only when \vgd exceeds a certain polarization voltage threshold VPTH. If \vgd is below the threshold VPTH, \fefet operates as a usual transistor and produces the current based on its polarization state.

\noindent \textbf{(2)} \fefets cannot be polarized above the level which the applied potential difference \vgd allows.

\noindent \textbf{(3)} Already polarized \fefets cannot be polarized further unless the applied potential difference \vgd is higher than the one this \fefet was polarized with previously.

\noindent \textbf{(4)} \fefet cannot be polarized indefinitely and can only achieve a certain polarization amplitude VPMAX regardless of the applied potential difference \vgd and its duration.

\noindent \textbf{(5)} Negative polarization occurs when the potential difference \vgd is negative and its absolute value exceeds VPTH. Polarization amplitude cannot go below the high-\vth state.

\noindent \textbf{(6)} Polarization rate depends on the value of the applied voltage difference \vgd.

\begin{table}
\centering
\caption{Polarization-related parameters in the \nn-based \fefet model.}
\label{table:pol_params}
\resizebox{\columnwidth}{!}{%
\begin{tabular}{l|l|l}
\toprule
\multicolumn{1}{c|}{Abbreviation} & \multicolumn{1}{c|}{Description} & \multicolumn{1}{c}{Default} \\ \midrule
VPMOD & Enables ($>$0) or disables polarization calculation & 1.0 \\
VPTH & Voltage threshold to initiate polarization & 1.5 \\
VPMAX &  Maximum achievable polarization amplitude & 4.5\\
A1-A5 &  Coefficients for polarization speed function & \cref{fig:halid_fit}\\
VPR &  Polarization rate coefficient & 1.0\\
VP0 &  Initial polarization level (normalized) & -1.0\\
VPDELTA &  Smoothing coefficient for gate/ceiling functions & 0.001\\
\bottomrule
\end{tabular}}
\end{table}

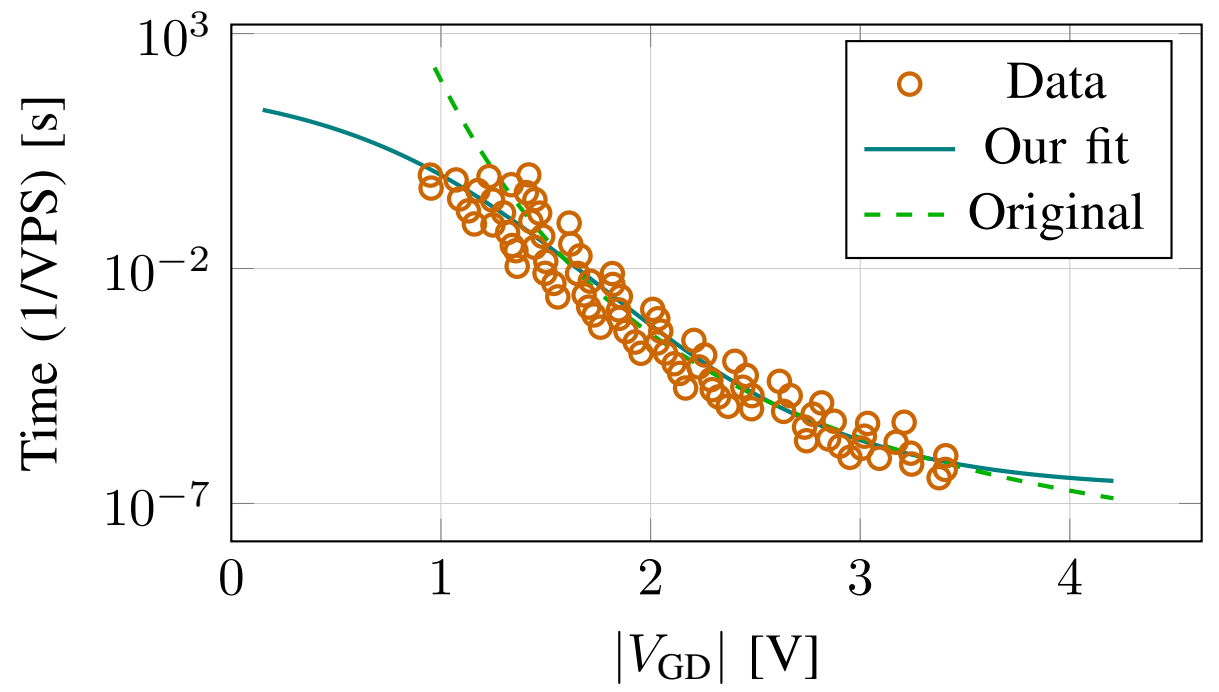
\begin{figure}[t]
    \centering
\begin{tikzpicture}
    \begin{axis}[
      ymode=log,
      xlabel={$|V_{\text{GD}}|$ [V]},
      xmin=0,
      ylabel={Time (1/VPS) [s]},
      grid=major,
      grid style={line width=.1pt, draw=gray!20},
      major grid style={line width=.2pt, draw=gray!40},
      legend pos=north east,
      width=8cm,
      height=5cm
    ]
      
      \addplot[
        only marks,
        mark=o,
          mark options={
            line width=0.8pt, 
            draw=orange!80!black,        
            fill=white,       
            scale=1.0         
          },
      ] table [col sep=comma, x index=0, y index=1] {1_initial/data/fit/halid_cloud.csv};
      \addlegendentry{Data}

      \addplot[
        thick,
        color=teal
      ] table [col sep=comma, x index=0, y index=1] {1_initial/data/fit/halid_fit.csv};
      \addlegendentry{Our fit}

      \addplot[
        thick,
        dashed,
        color=green!70!black
      ] table [col sep=comma, x index=0, y index=1] {1_initial/data/fit/halid_orig.csv};
      \addlegendentry{Original}

    \end{axis}
  \end{tikzpicture}
    \caption{Halid-plot of the proposed in \cref{eqn:pol_fun_3} polarization speed function for ERS data points of the \qty{28}{nm} \fefet \cite{Mulaosmanovic20} in comparison to the original in \cref{eqn:pol_fun_2}. Fitting parameters: VPR=1, A1=0.15, A2=10, A3=0.8, A4=1.4, A5=3.5.} 
    \label{fig:halid_fit}
\end{figure}

Statements from 1 to 5 are boundaries and conditions that need to operate in the time domain. Therefore, they are implemented using various gating functions combined into a differential equation. To derive a polarization speed VPS, we employ the formula for the \fefet switching time \cite{Mulaosmanovic20}:

\begin{equation}\label{eqn:pol_fun_2}
    \text{VPS} = \frac{d}{t_S} = \frac{d}{t_0} \exp \left[ -\frac{\alpha}{k_B T}\cdot \frac{1}{(V_{GD} - V_0)^2} \right]
\end{equation}

where $k_B$ is Boltzmann's constant, $T$ is the temperature, $t_0$, $\alpha$, and $V_0$ are the minimum switching time, exponential constant, and voltage offset, respectively. $d$ is the selected unit distance, which can be a unit of polarization amplitude or a normalized polarization unit. For simplicity, we assume that positive and negative polarization speeds have the same dependency with respect to the absolute value of \vgd. Since there is a denominator in \cref{eqn:pol_fun_2} which can be equal to zero when $V_{\text{GD}}$=$V_0$, this formula cannot be used in Verilog-A directly. Additionally, the function in \cref{eqn:pol_fun_2} is unbounded and may be prone to overflows during calculation. Therefore, we propose the following form for the speed function, which captures the dependency between polarization speed and \vgd:

\begin{equation}\label{eqn:pol_fun_3}
    \text{VPS} = \frac{\text{VPR}}{\text{A1} \cdot \exp \left[ \text{A2}\cdot \tanh (-\text{A3}\cdot |V_{GD}| + \text{A4}) - \text{A5} \right]} 
\end{equation}

The proposed function captures all necessary properties of the original function in \cref{eqn:pol_fun_2}: a non-linear exponential dependency of the VPS with respect to \vgd, asymptotically limited output when \vgd rises to infinity, or approaches $V_0$ and goes to 1/infinity. Infinite values cannot be simulated, but they can be limited to values small enough not to influence the simulation output. A1-A5 are function fitting coefficients responsible for the dependency between polarization speed and \vgd, and VPR is a polarization rate coefficient that can scale polarization speed. All polarization-related parameters can be found in \cref{table:pol_params}.

To demonstrate how the proposed in \cref{eqn:pol_fun_3} VPS function fits into data from the real device, we employ the timing data for a different \qty{28}{nm} \fefet \cite{Mulaosmanovic20}. We take the switching speed for ERS transitions and recalculate it into polarization speed using unit distance $d$=1. \cref{fig:halid_fit} demonstrates the resulting fit, which follows the data pattern while limiting the output values of the VPS function. Although there are some differences with the fit from the original function, the primary objective of the VPS function proposed in \cref{eqn:pol_fun_3} is to establish a simulation-compatible relationship between \vgd and polarization speed VPS.

\begin{figure*}[t]
    \centering
    \begin{subfigure}{\columnwidth}
        \phantomsubcaption
        \label{fig:iv_fit1}
        \phantomsubcaption
        \label{fig:iv_fit2}
        \phantomsubcaption
        \label{fig:iv_fit3}
    \end{subfigure}
    \input{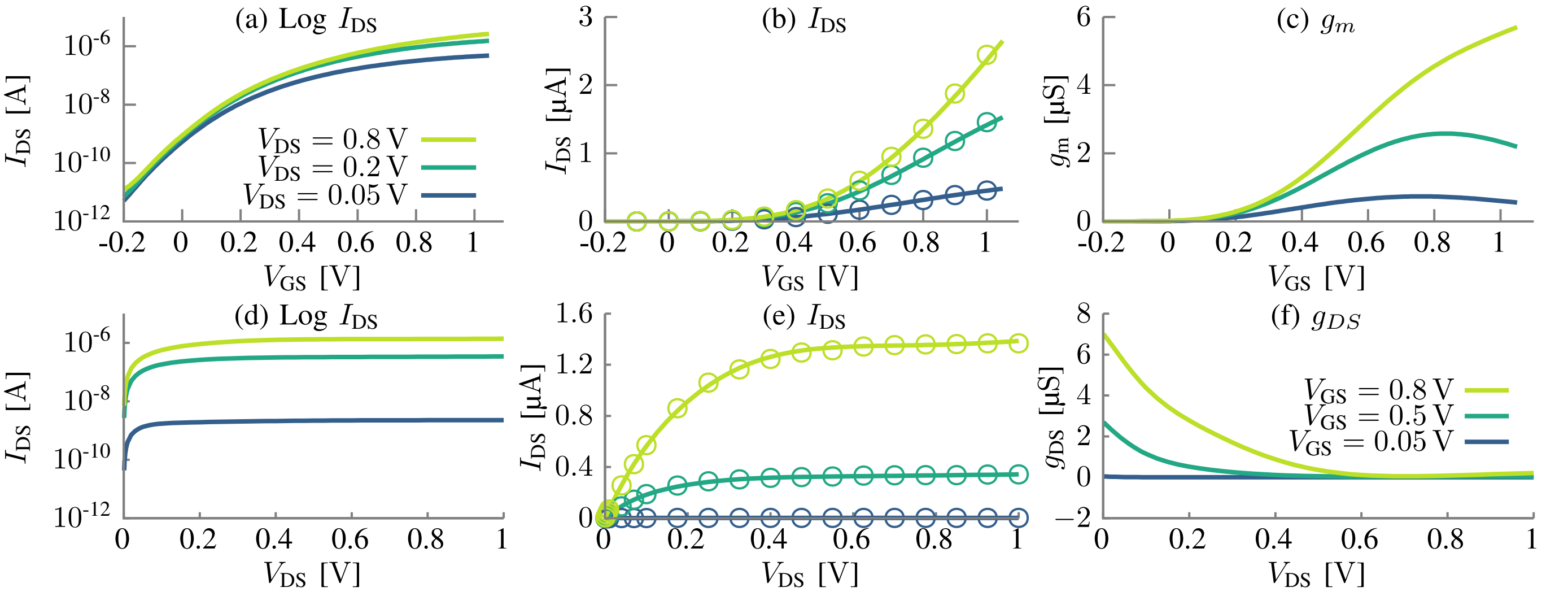}
    \vspace{-0.3cm}
    \caption{\idvg (a-c) and \idvd (d-f) fits for $\vpol=\qty{4.5}{V}$ demonstrate a good visual match with the measurement data while having smooth and stable derivatives. Lines are NN fit, circles are measurement data. Measurement data on logarithmic scale plots is not shown since it matches visually with the fit.}
    \label{fig:iv_fit}
\end{figure*}

\begin{figure}[t]
    \centering
    \begin{subfigure}{\columnwidth}
        \phantomsubcaption
        \label{fig:ivvp_fit1}
        \phantomsubcaption
        \label{fig:ivvp_fit2}
    \end{subfigure}
    \input{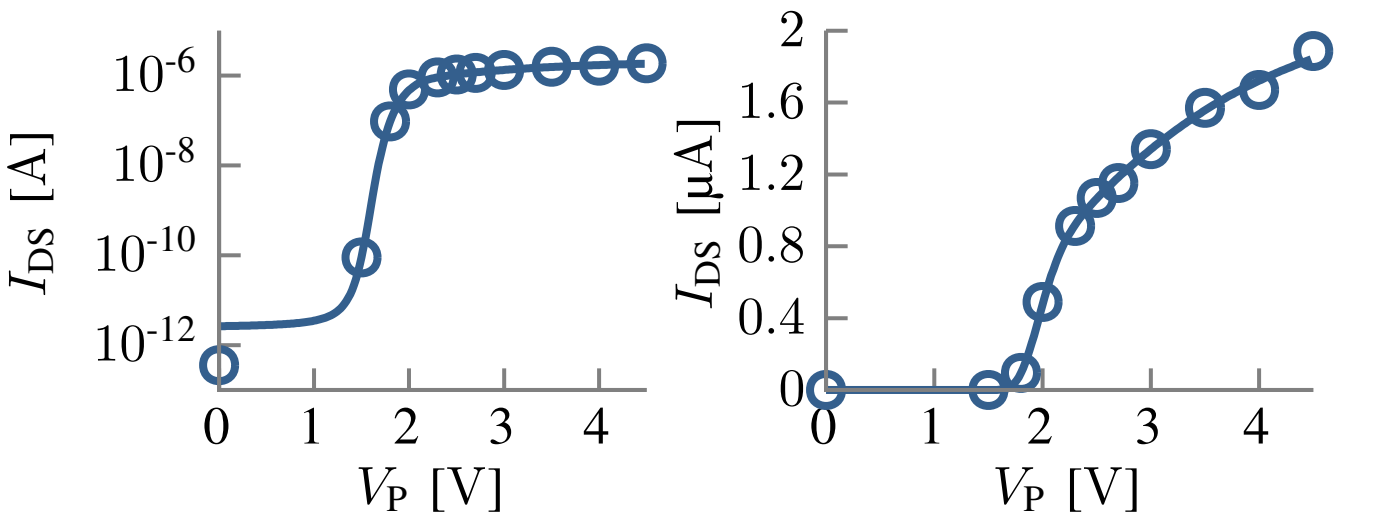}
    \vspace{-0.3cm}
    \caption{\ids-\vpol fit for $\vds=\vgs=\qty{0.9}{V}$ demonstrates a noticeable measurement drift for some polarization amplitudes (e.g. $\vpol=\qty{2.7}{V}$ and $\vpol=\qty{4.0}{V}$) that was corrected by \nn. Line is the \nn fit, circles -- measurement data.}
    \label{fig:ivvp_fit}
\end{figure}

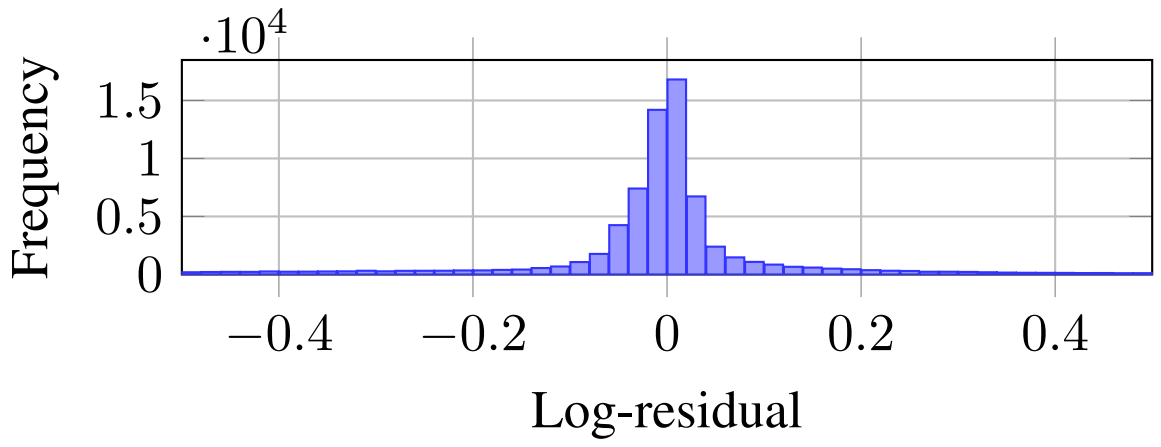
\begin{figure}[t]
  \centering
  \begin{tikzpicture}
    \begin{axis}[
        ybar,
        bar width=3.5pt,
        ymin=0,
        xmin=-0.5,
        xmax=0.5,
        xlabel={Log-residual},
        ylabel={Frequency},
        grid=major,
        height=3cm,
        width=8cm
    ]
      \addplot[fill=blue!40, draw=blue!80] 
        table[x=bin_center, y=count, col sep=comma] {1_initial/data/fit/binned_residuals.csv};
    \end{axis}
  \end{tikzpicture}
  \caption{Log-residuals between measured data and \nn fit resemble a normal distribution, which indicates that \nn fit filtered out noise present in measurement data and isolated the transistor's true physical behavior.}
  \label{fig:iv_dist}
\end{figure}

\section{Evaluation of \nn-based FeFET Model}
Since \nn was trained on noisy measurement data, any standard quantitative metric such as mean absolute error (MAE) or \mape will not yield meaningful results, as the final value will be heavily skewed by noise. Therefore, in addition to a regular visual evaluation, we also employ a statistical method to estimate the \nn fit quality. \cref{fig:iv_fit} illustrates \idvg (top) and \idvd (bottom) fits for \vpol=\qty{4.5}{V}.
Minor deviations between \nn fit and measured data are the result of the corrections that \nn has to perform with three-dimensional data.
Measurement data consists of \textit{independent} \idvg sweeps, which are coherent in one dimension only. However, \nn has to fit the data as a single smooth and continuous function $\ids=f(\vds, \vgs, \vpol)$ where all three dimensions are interconnected. Therefore, \ids values for an entire dimension or even a 2D plane may have to be shifted to achieve the required 3D fit. One of the prime examples for this behavior is the \ids-\vpol fit shown in \cref{fig:ivvp_fit}. During training, only 11 points shown in the figure for \vpol were used. Each of these 11 points is represented by a function $\ids^{\vpol}=f(\vds, \vgs)$. Thus, if measurements for one of \vpol points (see $\vpol=\qty{2.7}{V}$ or \qty{4.0}{V}) are noticeably off, the entire $\ids^{\vpol}$ plane has to be corrected for this particular \vpol, which inevitably leads to small mismatches between the \nn fit and measurements. A key advantage of forcing the \nn to learn a unified, continuous function is its effect on derivative quality. As illustrated in \cref{fig:iv_fit}, both the transconductance \gm and output conductance \gds are free from measurement noise artifacts and exhibit smooth and stable behavior.

In addition to a comprehensive visual evaluation, which includes all \idvg and \idvd sweeps for all polarization states, we also employ a statistical method to estimate the \nn fit correctness. A reliable way to estimate the fit is to evaluate the log-residuals between the measurement data and the \nn fit in all data points:
\begin{equation}\label{eqn:residuals}
    R = \log(\ids^{\text{meas}}) - \log(\ids^{\text{pred}})
\end{equation}

Despite the anisotropic structure of the raw measurement noise, \cref{fig:iv_dist} demonstrates that the resulting log-residuals approximate a zero-mean normal distribution. Thus, the residual error between the \nn predictions and the measurement data behaves as zero-mean Gaussian noise. This confirms that the \nn captures the underlying deterministic behavior for the drain current while filtering out measurement noise. By constraining the network's Lipschitz constant via spectral normalization, the model avoids overfitting to axis-aligned artifacts, effectively reducing complex measurement noise to uncorrelated Gaussian residual errors, achieving a smooth and unbiased fit.

\begin{figure*}[t]
    \centering
    \begin{subfigure}{\columnwidth}
        \phantomsubcaption
        \label{fig:memory_wr_in}
        \phantomsubcaption
        \label{fig:memory_wr_out}
    \end{subfigure}
    \input{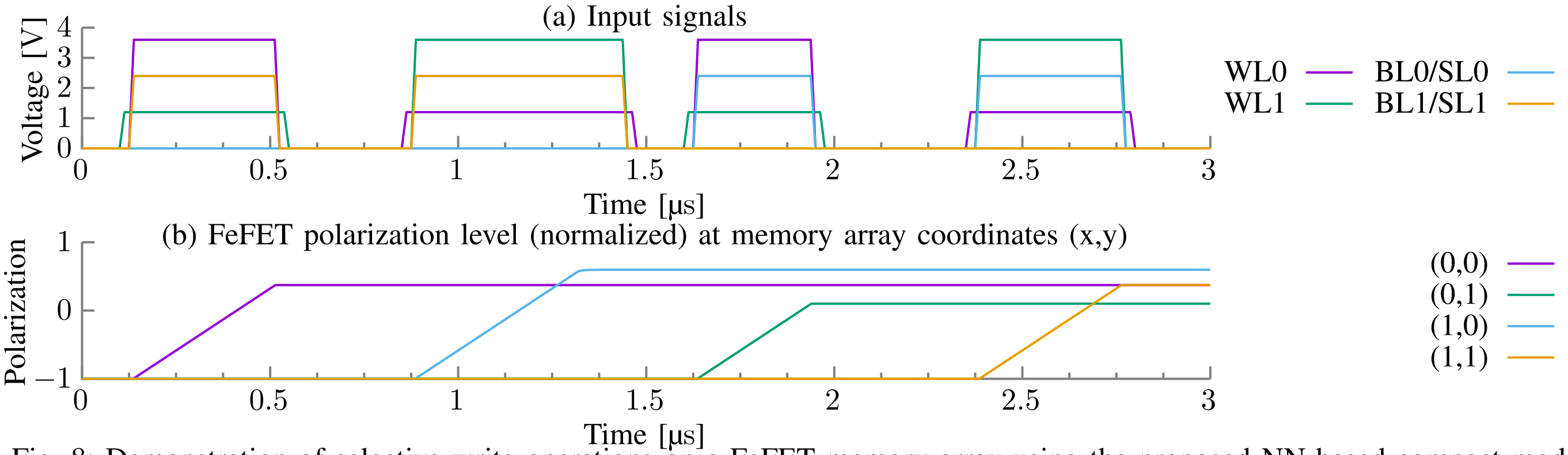}
    \caption{Demonstration of selective write operations on a \fefet memory array using the proposed \nn-based compact model. Memory array inputs (a) switch the polarization state of individual transistors (b). For clarity, a \grid{2} array configuration is depicted; however, the model was validated for array sizes up to \grid{512}.} 
    \label{fig:memory_wr}
\end{figure*}

\section{SPICE Simulations with \nn-based FeFET Model}
Although there are custom frameworks for performing SPICE-based \fefet array simulations \cite{park25}, all of our simulations were performed with a regular commercial SPICE simulator using our \nn-based Verilog-A \fefet model. We explore various \fefet memory arrays with different sizes and include parasitics represented by small resistance and conductance in each word, bit, and source lines for all \fefets. Therefore, an \grid{\text{M}} memory array netlist has $M^2$ transistors and $6\cdot M^2$ parasitic elements. The values for parasitic resistance and capacitance were set to \qty{0.5}{Ohm} and \qty{0.5}{fF}, respectively.

First, we simulate the writing scheme shown in \cref{fig:memory_write_scheme} using a \grid{2} \fefet memory array with $\vwrite=\qty{3.6}{V}$, which is divisible by three and ensures that $\sfrac{1}{3}\cdot \vwrite$ is below the polarization threshold voltage VPTH, meaning that polarization states of other cells will not be disturbed. The VPR for all \fefets is set to 4.5. VPR $=$ 4.5 in \cref{eqn:pol_fun_3} can be treated as if the speed was not calibrated based on a unit-distance but on a maximum polarization amplitude range. All other parameters remain at their default values as per \cref{table:pol_params}. \cref{fig:memory_wr} illustrates the writing scheme for all memory cells in the array. All \fefets start in the high-\vth state, and a specific set of voltage pulses is applied to the array at different time steps. The writing operation is performed as described in \cref{sec:memory_array_background} and all necessary signals can be found in \cref{fig:memory_wr_in}. The length of the pulse determines the polarization level. The first \vwrite pulse is applied to WL0 and targets the \fefet at array coordinates (0,0). The second set of voltage pulses lasts longer than the first one, resulting in a higher polarization state for \fefet at array coordinates (1,0). \fefet at array coordinates (0,1) exhibits the lowest level of polarization since the third set of pulses, which enables its writing, is shorter than both previous ones. The last set of voltage pulses is designed to replicate the first one. Therefore, the polarization state of \fefet at array coordinates (1,1) is the same as for the \fefet at array coordinates (0,0), demonstrating the stability of the polarization logic within the \nn-based transistor model.

\begin{figure*}[t]
    \centering
    \begin{subfigure}{\columnwidth}
        \phantomsubcaption
        \label{fig:memr_in}
        \phantomsubcaption
        \label{fig:memr_out}
    \end{subfigure}
    \input{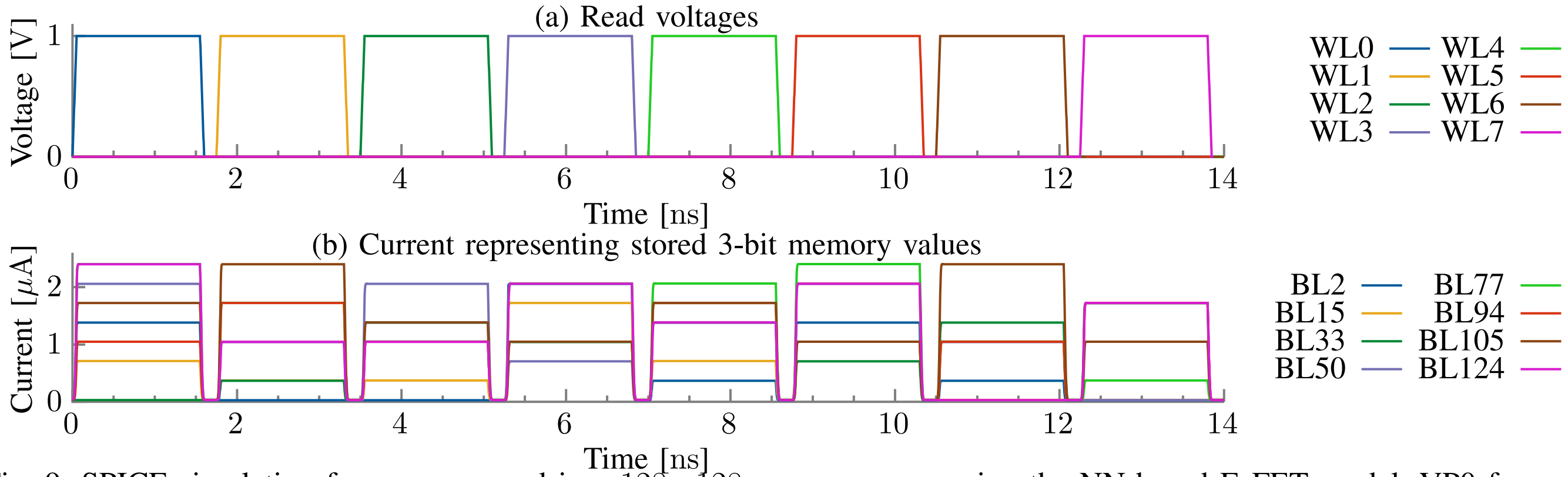}
    \caption{SPICE simulation for memory read in a \grid{128} memory array using the \nn-based FeFET model. VP0 for each transistor is initialized randomly from 8 states corresponding to a 3-bit memory. Each input (top) receives a read voltage pulse and produces an output in the form of a current (bottom).} 
    \label{fig:memr}
\end{figure*}

In contrast to the memory writing procedure, which primarily reflects the polarization logic correctness and behavior, memory reading evaluates the drain current calculation governed by the \nn within the \fefet model. To ensure a moderate level of simulation complexity, we employ a \grid{128} memory array with parasitics. For visual clarity, we simulate a 3-bit memory array with 8 states. First, we set the operating voltage to \qty{1}{V} and identify 8 polarization states corresponding to 8 equidistant drain currents in the \fefet with the fixed $\vds=\vgs=\qty{1}{V}$. Then, each \fefet in the \grid{128} memory array is assigned a random initial state VP0 from the calculated set of polarization states. \cref{fig:memr} illustrates the memory read simulation, where some selected input and output signals are shown. The read pulse of $\vread=\qty{1}{V}$ is applied to WL at different points in time, enabling the memory read for different rows of the array. Since constant voltage of \qty{1}{V} and \qty{0}{V} is applied to BL and SL, respectively, WL pulses in \cref{fig:memr_in} activate memory read in the first 8 rows of the array. Outputs in \cref{fig:memr_out} correspond to array columns which are selected randomly to cover the entire range of memory outputs. BL currents in \cref{fig:memr_out} have 8 clearly distinguishable levels which correspond to the selected states of a 3-bit memory. Although all levels are consistent across multiple bit lines, the transition behavior slightly differs across these lines, which is not visible due to scale. The difference in transition behavior is caused by parasitic elements in the array, which slow down the transition. In particular, bit lines far apart in the array layout will exhibit the most difference in transition behavior due to a large number of accumulated parasitic elements.

\begin{figure}
    \centering
    \begin{tikzpicture}
    \begin{axis}[
        width=\columnwidth,
        height=3.7cm,
        xlabel={Memory array size (M)},
        ylabel={Simulation time [s]},
        xmode=log,
        ymode=log,
        log basis x={2},
        log basis y={10},
        grid=both,
        grid style={dashed, gray!50},
        minor grid style={dashed, gray!30},
        xtick={8,16,32,64,128,256,512},
        xticklabels={8,16,32,64,128,256,512},
        ymin=0.5,
        ymax=10000,
        legend pos=north west,
        mark size=3pt,
        every node near coord/.append style={
            font=\footnotesize,
            anchor=south,
            shift={(5pt,3pt)}
        },
    ]
    
    \addplot[
        color=blue,
        mark=*,
        nodes near coords,
        point meta=explicit symbolic,
        visualization depends on={y \as \myy},
        every node near coord/.append style={
            /pgf/number format/fixed,
            /pgf/number format/precision=1
        },
    ] coordinates {
        (8, 0.8)   [0.8]
        (16, 1.8)  [1.8]
        (32, 4.0)  [4.0]
        (64, 17) [17]
        (128, 66)  [66]
        (256, 229) [229]
        (512, 1020)[886]
    };
    \addlegendentry{VPMOD on}

    \addplot[
        color=red,
        mark=*,
        nodes near coords,
        point meta=explicit symbolic,
        visualization depends on={y \as \myy},
        every node near coord/.append style={
            anchor=north,
            yshift=-5pt,
            /pgf/number format/fixed,
            /pgf/number format/precision=1
        },
    ] coordinates {
        (128, 57)  [57]
        (256, 210) [210]
        (512, 737)[737]
    };
    \addlegendentry{VPMOD off}
    
    \end{axis}
    \end{tikzpicture}
    \caption{SPICE simulation time for \grid{\text{M}} memory arrays with the Verilog-A \nn-based FeFET model. Polarization re-calculation is enabled for blue and disabled for red. Polarization state VP0 is initialized randomly for each transistor.} 
    \label{fig:memory_time}
\end{figure}
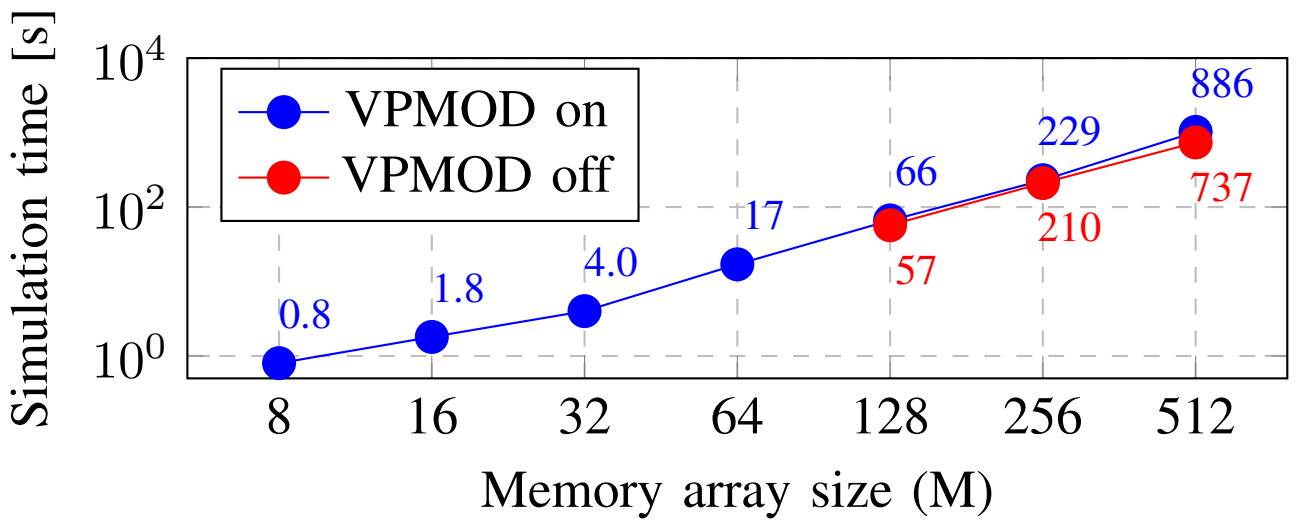

Another crucial aspect of the transistor model performance is the simulation time. For simulation time measurements, we employ the memory array architecture from previous experiments and vary its dimensions. All parasitic elements are also included to increase the simulation's complexity. The general setup for the experiment and input signals is very similar to that of the previous simulation, with the \grid{128} array shown in \cref{fig:memr}. In contrast to the previous experiment, VP0 was not selected from a 3-bit set but was initialized randomly with values in the range [-1, 1]. The transient simulation time is fixed at \qty{15}{ns} across all simulations to equalize conditions. \cref{fig:memory_time} illustrates how the size of the memory array affects SPICE simulation time with the \nn-based \fefet model. Since we employ square \grid{\text{M}} memory arrays, the number of transistors increases quadratically, and the simulation time increases exponentially. All simulations converged without any issues with the same simulation options across all array sizes. A \grid{512} memory array with \num{262144} \fefets and more than 1.5 million parasitic elements takes around 15 minutes to simulate. Disabling polarization logic in the model via a VPMOD switch reduces simulation time for large memory arrays by up to 17\%.

\section{Conclusion} \label{sec:conclusion}
In this work, we presented a derivative-free neural network \nn-based modeling approach aimed to accelerate early-stage circuit prototyping for emerging transistor technologies using only measurement data. By incorporating spectral normalization, our method effectively mitigates complex measurement noise and ensures smooth derivative behavior without relying on numerical solver gradients during training. We validated our approach on a \qty{28}{nm} \fefet, accurately capturing a continuous range of operating voltage conditions and polarization amplitudes. Furthermore, the model's scalability and compatibility with standard SPICE engines were demonstrated through transient simulations of large-scale memory arrays with up to \grid{512} size and parasitic effects. By bypassing the prolonged, expert-intensive process of traditional transistor model development, our approach provides a fast and reliable path from raw measurement data to circuit simulation, significantly accelerating the deployment cycle for next-generation semiconductor devices.

\section*{Acknowledgment}
We would like to thank Halid Mulaosmanovic, Stefan Dünkel, and Sven Beyer from GlobalFoundries for their valuable support and for providing the FeFET wafers for measurements.

\end{document}